\documentclass[usegraphicx,usenatbib,a4paper]{mn2e}
\usepackage{aas_macros,graphicx,times,multirow}
\usepackage{amssymb}
\usepackage{float}

\title[Swift J2218.4+1925, a new hard Polar]
{\J22\ : a new hard X-ray selected Polar observed with \XMM}
\author[F. Bernardini et al.]                                                    
{F.~Bernardini,$^{1,2}$\thanks{E-mail: bernardini@wayne.edu} 
D.~de Martino$^{2}$, 
K.~Mukai$^{3,4}$, 
M.~Falanga$^{5,6}$\\
$^1$ Wayne State University, 666 W. Hancock Street, Detroit, MI, USA\\
$^2$ INAF $-$ Osservatorio Astronomico di Capodimonte, Salita Moiariello 16, I-80131 Napoli, Italy\\
$^3$  CRESST and X-Ray Astrophysics Laboratory, NASA Goddard Space Flight  Center, Greenbelt, MD 20771, USA\\
$^4$ Department of Physics, University of Maryland, Baltimore County, 1000 Hilltop Circle,Baltimore, MD 21250, USA\\
$^5$ International Space Science Institute (ISSI), Hallerstrasse 6,CH-3012 Bern, Switzerland\\
$^6$ International Space Science Institute in Beijing, No. 1 Nan Er Tiao, Zhong Guan Cun, Beijing 100190, China}

\date{}
\pagerange{\pageref{firstpage}--\pageref{lastpage}}
\def\Swift{{\em Swift}}

\def\XMM{{\em XMM-Newton}}

\def\ergscm{$\rm erg\,cm^{-2}\,s^{-1}$}
\def\INT{{\em INTEGRAL}\,}

\def\J22{Swift\,J2218.4+1925}

\begin{document}

\label{firstpage}

\maketitle

\begin{abstract}

\J22, a hard X-ray source detected by \Swift\ BAT, 
has been proposed as a candidate magnetic cataclysmic variable of the polar type from optical spectroscopy. 
Using \XMM\ we perform detailed timing and spectral analysis with simultaneous 
X-ray ($0.3-10$ keV) and optical B band data. We complement the spectral study with 
archival hard X-ray ($14-70$ keV) spectra collected 
by \Swift\ BAT as well as with optical, near
and mid-infrared photometry from $SDSS$, $2MASS$ and $WISE$ archive, respectively.
A strong periodic X-ray signal at 2.16 h, fully consistent with the 
recently determined spectroscopic orbital period, adds  
\J22\  to the small group of hard X-ray polars and locates it at the low edge of the orbital period gap.  
The X-ray pulse profile shows the typical bright and 
faint phases seen in polars, that last $\sim 70\%$ and $\sim 30\%$ of the orbit, 
respectively. A pronounced dip centred on the bright phase is also detected. 
It  is stronger at lower energies and is mainly produced by photoelectric absorption. 
A binary inclination $i \sim 40^o - 50^o$ and a magnetic colatitude 
$\beta \sim 55^o - 64^o$ are estimated. 
The source appears to accrete over a large area $\sim 24^o$ wide.
A multi-temperature optically thin emission with complex absorption
well describes the broad-band ($0.3-70$ keV) spectrum, with no signs 
of a soft X-ray blackbody component. The spectral shape strongly varies with the 
source rotation reaching plasma temperatures up to 55\,keV,   
hardening at the dip and being softer during the faint phase ($\sim7$ keV).
We also find the first indication of 
an absorption edge due to a warm absorber in a polar. 
Indication of overabundance of neon is found in the RGS spectra.
The UV to mid-IR spectral energy distribution reveals an excess in the near and mid-IR, 
likely due to low cyclotron harmonics.
We further estimate a WD mass of 0.97 M$_{\odot}$ and a distance of $230-250$ pc. 
\end{abstract}


\section{Introduction}


 Cataclysmic Variable stars (CVs) are X-ray binaries hosting a white dwarf (WD) 
that is accreting matter from a companion star (a late-type main-sequence or a sub-giant) 
with mass smaller than that of the Sun.
 The \Swift\ Burst Alert Telescope \citep[BAT,][]{Barthelmy} and 
 and \INT\ IBIS \citep{ubertini03} surveys detected a non-negligible number of new CVs 
\citep{Cusumano10,Baumgartner13,bird10}. Most of the known hard X-ray detected CVs are magnetic 
 CVs of the intermediate polar type \citep{barlow06}, a few are instead polars.
The first host asynchronously ($P_{rot}<P_{orb}$) rotating WDs with possibly weaker magnetic 
fields (B$\leq10^{6}$ G), while the latter contain synchronously ($P_{rot}=P_{orb}$) 
rotating WDs with stronger magnetic fields, B$\sim 10-230\times 10^{6}$ G \citep[see also][for 
a review]{Warner95}. 
The incidence of magnetism in the hard X-ray selected samples is very high, 
up to $\sim 90\%$. Magnetic CVs, especially the intermediate polars, 
are debated to represent the main constituent of the  
galactic population of X-ray sources at low luminosities \citep{muno04,sazonov06,revnivtsev09,revnivtsev11}.
Newly discovered sources identified through optical photometry 
\citep[e.g.][]{Masetti08,Masetti10,Masetti12a,Masetti12b} have added as candidates
to the group of magnetic CVs. 
However, the proper identification and classification of new sources relies on X-rays,
as successfully demonstrated in the recent years \citep[e.g.][]{demartino08,Anzolin08,Anzolin09,Bernardini12,Bernardini13}, that indeed reveal that
most of the hard X-ray CV candidates are of the magnetic type. We here present the
results of a dedicated \XMM\ pointing to assess the true nature of the new hard X-ray selected source, \J22.

 \J22\ was detected by the BAT instrument onboard \Swift\  and is coincident with 
1RXS\,J221832.8+192527. Using  optical time-resolved spectroscopy, \citet{Thorstensen13} derived an 
orbital period of $7773\pm7$ s.
 The broad wings and the large amplitude of radial velocities of the H$_{\alpha}$ emission line 
 also suggested a magnetic nature for this CV. The additional presence of a narrow 
 component in the emission line profiles, antiphased with respect to the broad 
component, and ascribed to the irradiated face of the secondary star, further suggested a Polar-type magnetic CV. 
However, the detection of  X-ray pulsations at 
a period consistent with that of orbital spectroscopy, as well as optical polarimetry are mandatory to 
confirm the classification. 
Up to now \J22\ was observed in the soft X-ray range ($<10$ keV) only sporadically with the XRT 
instrument on board \Swift\, and no X-ray periodicity was found \citep{Thorstensen13}. 
\XMM\ has proven to be the best observatory, currently operating, to identify periodic signals
on the timescales exhibited by CVs (minutes - hours) and to characterize their spectral properties 
\citep[see e.g.][]{Bernardini12,Bernardini13}. 
In Section \ref{sec:obs} we present the \XMM\ data complemented with the \Swift\ BAT data. 
In Section \ref{sec:results} we presents the X-ray results together with optical, 
 near and mid-infrared photometry from the Sloan Digital Sky Survey ($SDSS$), the  Two-Micron All Sky Survey 
($2MASS$)  and the Wide-field Infrared Survey Explorer \citep[\emph{$WISE$,}][]{Wright10}. The results are
discussed in Section \ref{sec:discuss}, in terms of emission components.


\section{Observations and data analysis}
\label{sec:obs}

\subsection{\textit{XMM-Newton} observations}

\J22\ was observed in November 2013 by \XMM\ observatory \citep{turner01,mason01,denherder01} 
with the EPIC cameras (PN, MOS1, and MOS2) as main instruments. 
The details of the observation, together with that of \Swift\, 
are reported in Table \ref{tab:observ}. The \XMM\ data were processed using 
the \textsc{SAS} version 13.0.0 and the latest calibration files (CCF) available
on October 2013.

\subsubsection{The EPIC and RGS data}

The three EPIC cameras were all set in Prime Full Window imaging mode with the thin filter applied.
Standard data screening criteria were applied for all instruments. 
For the EPIC data we extracted the source photons from a circular region of radius 37.5 arcsec centered at the source position. 
Background photons were taken from a nearby region of the sky of radius 70 arcsec, 
clear from source contamination, in the same CCD where the source lies.
For the spectral analysis, in order to avoid background solar flare contamination, we generated the source and the 
background spectra removing high background epochs in all three 
instruments. 
For the timing analysis, instead, we used the whole data set. 
We also produced background-subtracted light curves in the ranges 
$0.3-12$ keV (with a bin time of 15 s), 
$0.3-1$ keV, $1-3$ keV, $3-5$ keV and $5-12$ keV (with a bin time of 75 s).  
The event arrival times were corrected to Solar system barycenter by using the task \textsc{barycen}. 
The average (whole observation) EPIC spectra were rebinned before fitting in order to have a minimum of 25 counts each bin. 
Spectra were also extracted as a function of the source rotational phase (pulse-phase spectroscopy analysis).
We report the spectral results of the simultaneous analysis of the three cameras in order to increase the signal to noise (S/N). The spectral fit were made with \textsc{Xspec} version 12.7.1.

The two RGS instruments were operated in Spectroscopy mode. The source
is quite faint, with the net count rates reported in Table \ref{tab:observ}
corresponding to 56\% and 67\% of the total (source plus background) count
rates for RGS1 and RGS2, respectively. 

\begin{table*}
\caption{Summary of main observations parameters for all instruments. Uncertainties are at $1\sigma$ confidence level.}
\begin{center}
\begin{tabular}{ccccccc}
\hline
  & & & & & & \\
 
Telescope      & OBSID & Instrument & Date                    & UT$_{\rm start}$ & T$_{\rm expo}^{a}$  & Net Source Count Rate\\
                                      &              &        & yyyy-mm-dd      & hh:mm & (ks)      &    c/s                  \\
\hline
 \emph{XMM-Newton}&  0721790101 & EPIC-pn   & 2013-11-25  & 19:47 & 31.5    & $1.263\pm0.007$ \\
                      &              & EPIC-MOS1 & 2013-11-25  & 19:24 & 33.1 & $0.359\pm0.03$     \\ 
                      &              & EPIC-MOS2 & 2013-11-25  & 19:24 & 33.1 & $0.361\pm0.03$     \\  
                      &              & RGS1      & 2013-11-25  & 19:23 & 33.4 &$0.0184\pm0.0012^b$ \\
                      &              & RGS2      & 2013-11-25  & 19:23 & 33.4 &$0.0257\pm0.0012^b$ \\
                      &              & OM-B      & 2013-11-25  & 19:30 & 28.0 & $3.85\pm0.02$ \\
      \emph{Swift} & $^{c}$          & BAT &  &	& $1\times10^{7}$ & $1.6\pm0.3\times 10^{-5}$ \\ 
\hline
\end{tabular}
\label{tab:observ}
\end{center}
\begin{flushleft}
$^{a}$ Net exposure times.\\
$^{b}$ We report the net count rate in the 0.4--2.0 keV band, excluding dead channels. \\
$^{c}$ All available pointings collected during 2004 December to 2010 September are summed together.\\
\end{flushleft}
\end{table*}

\subsubsection{The optical monitor photometry}

The Optical Monitor (OM) instrument, operated in fast window mode,  
observed \J22\ with the B filter, centred at 4500\,$\rm \AA$, simultaneously to the EPIC cameras. 
A total of 10 series of $\sim$2800\,s were obtained (see Table \ref{tab:observ}).
\J22\ is found at an average magnitude B=17.8$\pm$0.05 corresponding to a flux
$\rm 4.80\pm0.06\times10^{-16}\,erg\,cm^{-2}\,s^{-1}\,\AA ^{-1}$. It is at similar flux
level as when detected in the Sloan Digital Sky Survey ($SDSS$), $g'$=17.64$\pm$0.01 and
when observed by \citet{Thorstensen13}, both in 2009.  
The OM light curves as obtained from the standard processing pipeline were also corrected to the solar system barycentre. 

\subsection{The \Swift\ observations}

BAT is sensitive in the 14-195 keV energy range and 
has built up a all-sky map of the hard X-ray sky thanks to its wide field of view. 
We downloaded the eight-channel spectra from the first 70 month of monitoring directly 
available at http://swift.gsfc.nasa.gov/results/bs70mon/.

\section{Results}
\label{sec:results}

\subsection{The X-ray timing analysis}
\label{subsub:timing}

\begin{figure}
\centering
\includegraphics[angle=270,width=3.3in]{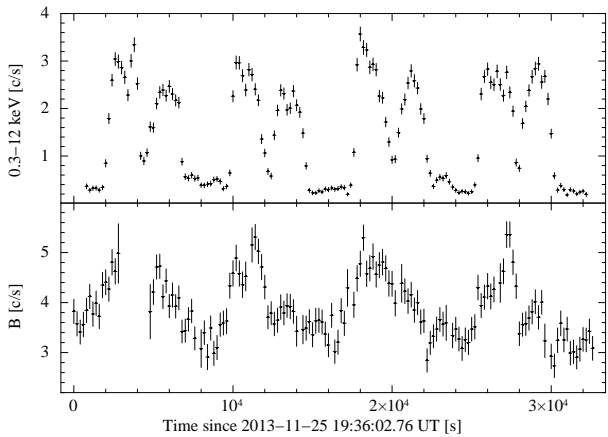}
\caption{\textit{Upper panel}: 0.3-12 keV background subtracted PN light curve. \textit{Lower panel}: Optical B band light curve. The binning time is 200 s.}
\label{fig:xb_lc}
\end{figure}

The $0.3-12$ keV background subtracted PN light curve is shown in Fig. \ref{fig:xb_lc}. 
A periodic modulation is evident (the observation covers four cycles), 
showing a bright double-humped pulse.
The power spectrum in the $0.3-12$ keV range reveals a strong peak at 
$\sim1.3\times10^{-4}$ Hz together with harmonics up to the fourth.
A period $\rm P=7770\pm10$ was determined by fitting the PN $0.3-12$ keV light curve 
with a composite sinusoidal function with the fundamental 
frequency plus its first four harmonics. 
This is consistent within $1\sigma$ with the period derived from optical spectroscopy 
\citep[$7773\pm7$ s][]{Thorstensen13}.
All uncertainties are hereafter at $1\sigma$ confidence level if not otherwise specified.

\begin{figure}
\centering
\includegraphics[angle=-90,width=4.3in]{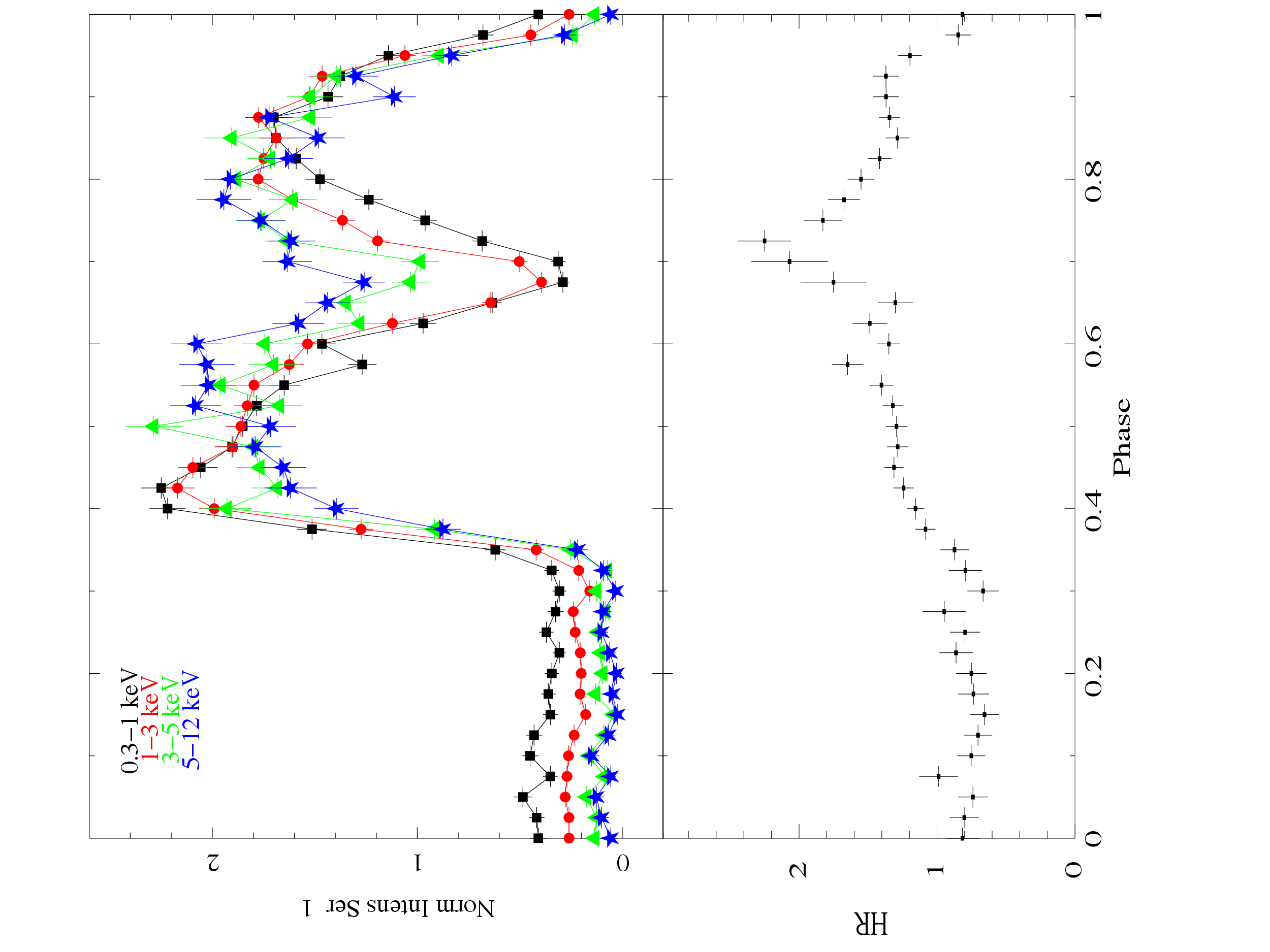}
\caption{\textit{Upper panel}: PN background subtracted light curves, folded at $P=7770$ s, in four energy intervals: $0.3-1$ (black squares), $1-3$ (red circles), $3-5$ (green triangles), $5-12$ (blue stars) keV. 
\textit{Lower panel}: Hardness ratio, 1--3 keV vs 0.3--1 keV.}
\label{fig:pulsvse}
\end{figure}

\noindent We folded the background-subtracted light curves at $P=7770$ s. 
They show a relatively long ($70\%$) bright phase
and a faint phase ($\sim 30\%$) where the countrate goes close to zero only for E$>3$ keV. 
The bright phase is characterised by a double-humped maximum 
with a slightly stronger first peak and a prominent dip between the two (see Fig. 
\ref{fig:pulsvse}). The pulse profile structure slightly evolves with energy, 
 and the dip is clearly more pronounced at lower energies. 
To measure the fractional intensity of the dip ($\Delta\,I$)
we fit the pulse bright phase in the range 0.4--0.9 with a linear 
function, accounting for the overall decay in the count rate in this range,
plus a Gaussian accounting for the dip. We left the intensity of the Gaussian ($I$), 
the semi-half width at half intensity ($\sigma$) and the phase centre ($\phi_{cent}$) 
free to vary. We then define the fractional intensity of the dip: 
$\Delta\,I = I/I_{cent}$, where $I_{cent}$ is the interpolated continuum count rate 
evaluated at the dip centre. We find: 
$\Delta\,I$ is $85\pm8\%$ ($0.3-1$ keV), $76\pm6\%$ ($1-3$ keV), 
$38\pm5\%$ ($3-5$ keV), and  $12\pm4\%$ ($5-12$ keV). 

We also inspected the times of dip occurrence during the four cycles covered by the 
\XMM\ observation (see Fig. \ref{fig:xb_lc}). The first and third dip occur at similar phase while the  second is leading 
by $30\pm15$ s and the fourth is delayed by $110\pm15$ s. This is not evidence of period change, but 
rather a signature of changes or oscillation in the accretion flow rate. 
This behaviour is not uncommon in polars \citep{HarropAllin99,Schwope01}.
 
\noindent We also calculated the pulsed fraction (PF) of the fundamental harmonic here defined as: $\rm PF=(A_{max}-A_{min})/(A_{max}+A_{min})$. $\rm A_{max}$ and $\rm A_{min}$ are the maximum and minimum value of the sinusoid used to 
fit the fundamental harmonic. The PF increases with the energy interval being 
PF$_{\rm 0.3-1 keV}=49\pm1\%$, PF$_{\rm 1-3 keV}=68.9\pm0.8\%$, PF$_{\rm 3-5 keV}=95\pm1\%$, 
PF$_{\rm 5-12 keV}=102\pm1\%$. We also inspected  the spectral behaviour versus phase measuring the hardness ratio (HR) 
defined as $\rm HR=n^{\phi}_{keV^{a}}/n^{\phi}_{keV^{b}}$, where $\rm n^{\phi}_{keV}$ is the number of photons 
in the inspected energy range (called a and b), and \textit{$\phi$} is the phase interval. 
We found a strong hardening at the phase corresponding to the dip ($\phi\sim0.7$)  
for both $\rm HR=n^{\phi}_{1-3}/n^{\phi}_{0.3-1}$ and $\rm HR=n^{\phi}_{3-5}/n^{\phi}_{1-3}$ 
(see Fig. \ref{fig:pulsvse}).

\subsection{The optical timing analysis}
\label{subsub:optical}

\begin{figure}
\centering
\includegraphics[angle=270,width=3.3in]{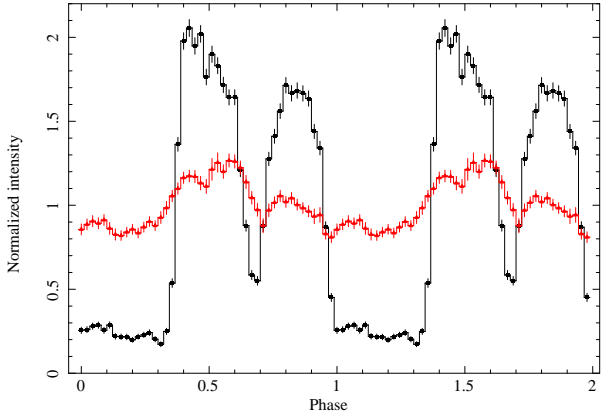}
\caption{X-ray ($0.3-12$ keV, black circles) vs optical (B, red triangles) folded light curves. The folding period is 7770 s. 
The two folded light curves are over-plotted to allow direct comparison. 
Two cycles are shown for plotting purposes.}
\label{fig:x_vx_b_pulse}
\end{figure}

A fit made with a single sinusoid to the B band light curve 
gives a period of 7834$\pm$44 s, consistent within 1.5$\sigma$ 
with that found from X-rays. The shape of the pulse profile is similar to that in the X-rays,
but with a slower rise to the maximum and a much weaker secondary maximum.
The intensity of the pulsation, PF$_{\rm B}=16.8\pm0.7\%$, is much lower than in the 
$0.3-12$ keV band, PF$_{\rm 0.3-12 keV}=79.2\pm0.5\%$ (see Fig. \ref{fig:x_vx_b_pulse}). 
A cross-correlation between X-ray ($0.3-12$ keV) and B band light curves gives the optical 
leading the X-ray light by $\sim$700 s. This is mainly dictated by the early rise to the maximum
with respect to the X-rays. 
 
\noindent  The B band pulse also reveals an optical 
counterpart of the X-ray dip that is 
also occurring at $\phi\sim0.7$, but with a slower decay. 
We found $\Delta\,I_{B}=18\pm3\%$, which is similar to that of the $5-12$ keV range ($\Delta\,I=12\pm4\%$), but much lower than that in the $0.3-1$ keV range ($\Delta\,I=85\pm8\%$). 
The dip has a width of about 0.1 in phase, both in the X-ray and in the B band. 

\subsection{X-ray spectral analysis}
\label{sub:spec}

\begin{figure}
\centering
\includegraphics[angle=0,width=3.5in]{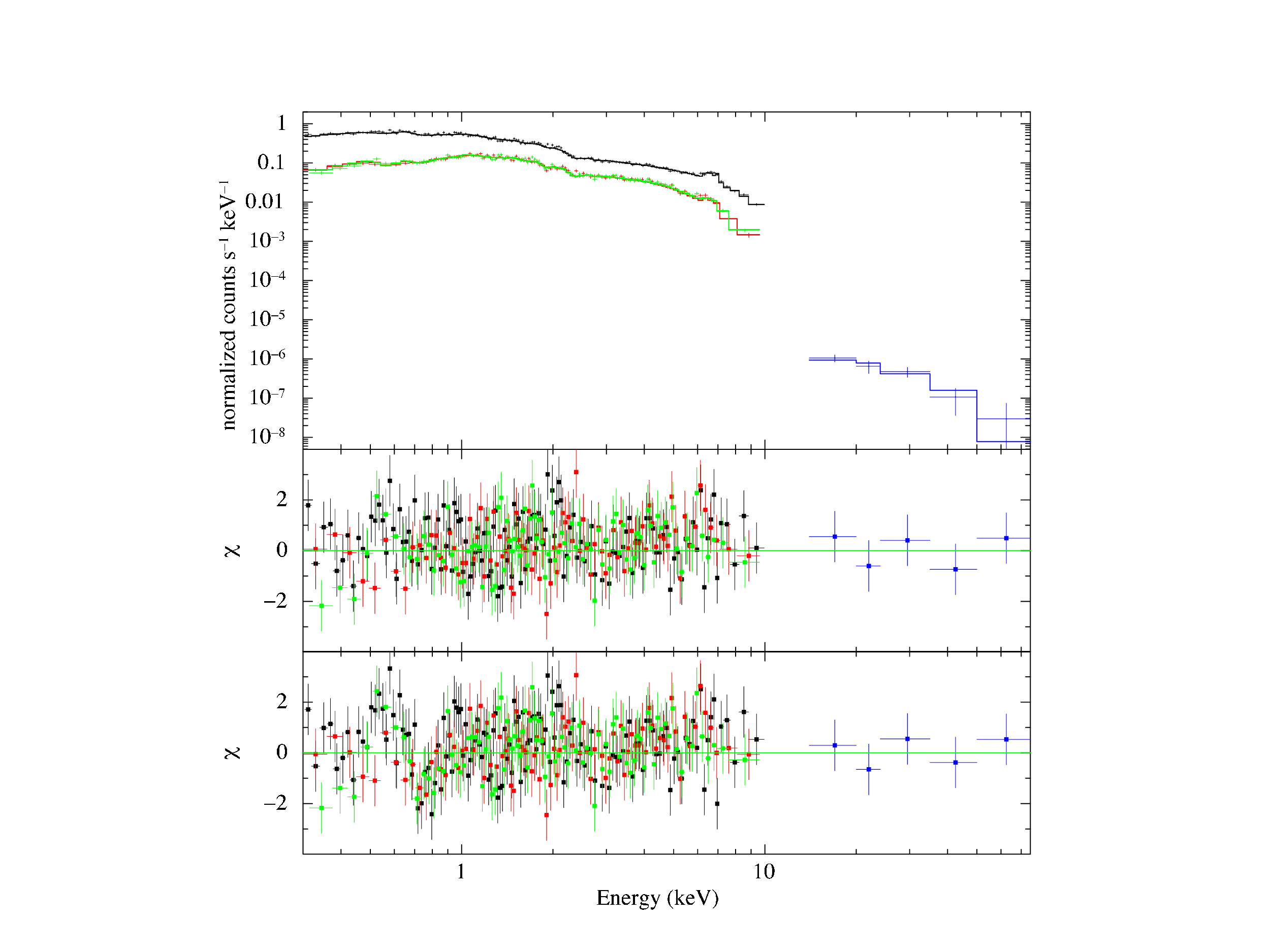}
\caption{\textit{Upper panel:} Broad band $0.3-100$ keV spectrum of \J22.
The PN spectrum is shown in black, the MOS1 and MOS2 spectra are in red and green respectively, 
while the BAT spectrum is in blue. The continuum lines represent the best-fitting model.
\textit{Central panel:} Post fit residuals with respect to the best-fitting model including the \textsc{edge}. 
\textit{Lower panel:} Post fit residuals without the inclusion of the \textsc{edge}.}
\label{fig:avspec}
\end{figure}

We simultaneously fitted the average spectrum of the three EPIC cameras together with that of BAT.
A single temperature model is inadequate to fit the resulting broad band ($0.3-70$ keV) spectrum. 
The best fitting model is a combination of a multi-temperature-like optically thin plasma 
(\textsc{cemekl}) plus a Gaussian at 6.4 keV (emission line), including a total (\textsc{phabs}) and 
a partial covering (\textsc{pcfabs}) absorber. They  
account for the absorptions from the galactic interstellar medium and from a local, 
partial covering, cool material close to the source. We linked all model parameters among 
different instruments, 
leaving free to vary only a normalisation parameter accounting for a possible difference in the calibration. 
The \textsc{cemekl} model is normally used to account for a gradient of temperature in the post-shock region of CVs.
The emission measure follows a power law in temperature $dEM=(T/T_{max})^{a-1}\,dT/T_{\rm max}$. 
We obtained kT$_{\rm max}=37\pm7$ keV and $\alpha=1.1\pm0.1$. 
The total absorber column density, N$_{H_{\rm P}}=3.3\pm0.3\times10^{20}$ cm$^{-2}$, is lower than the 
galactic value in the direction of the source ($4.4-4.8\times10^{20}$ cm$^{-2}$) derived from \cite{Kalberla05} 
and \cite{Dickey90}, implying an origin in the galactic interstellar medium.
On the other hand, the partial covering absorption component has 
N$_{H_{\rm pc}}=4.8\pm0.5\times10^{22}$ cm$^{-2}$ and a covering fraction $cvf=42\pm2\%$, implying 
a localized origin. The fluorescent Fe 6.4 keV emission line has an equivalent width $EW=0.10\pm0.02$ keV.
We also noticed in the fit residual the presence of a feature in all three instruments (see Fig. \ref{fig:avspec} lower panel) and we modeled it with an \textsc{edge}. 
The \textsc{edge} parameters where linked between different instruments, but left free 
to vary to minimize the $\chi^2$. We measure $E_{edge}=0.69\pm0.01$ keV, and $\tau_{edge}=0.35\pm0.06$, 
where $E_{edge}$ is the threshold energy and $\tau_{edge}$ the absorption depth at the threshold.
The fit has $\chi^2=1324.63$ for 1299 degrees of freedom (dof), 
compared to $\chi^2=1371.08$ for 1301 dof, for the fit without the \textsc{edge}.
All spectral parameters are reported in 
Table. \ref{tab:avspec}, and the average spectrum is showed in Fig. \ref{fig:avspec}.
We stress that the feature is simultaneously detected in the PN, MOS1 and MOS2 spectra 
and that $\Delta\chi^2$ is quite high (46.45). 
This points toward a real detection more than to a feature produced 
by a random statistical fluctuation. 

\noindent In order to properly estimate the significance of the absorption feature we run extensive Monte Carlo 
simulation by producing $10^{4}$ spectra for each instrument (PN, MOS1, and MOS2)  
using as underling continuum the best fit model used for the average spectrum, without the \textsc{edge}  
\citep[see e.g.][for more details on the technique]{lanzuisi13,bernardini09}. 
The simulated spectra have the same continuum flux, background flux and exposure as observed. 
We first fitted the simulated spectra with the best fitting model without the \textsc{edge}, 
and then with the same model including the \textsc{edge}. 
The \textsc{edge} energy was left free to vary in the full 0.3--10 keV energy range. Also 
$\tau_{edge}$ was left free to minimize the $\chi^2$. 
Then, we counted how many times we obtained a $\Delta\chi^2$ greater than the observed one (46.45), 
just because of statistical fluctuations. Since we never found 
a $\Delta\chi^2$ greater than the observed one,
this implies that we found 0 spectra over $10^{4}$ 
set of PN and MOSs spectra presenting such intense absorption feature 
just because of statistical fluctuations. This leads to an 
estimated significance level greater than $3.9\sigma$ for the 0.7 keV edge.

\noindent A blackbody component with kT$\sim$20-60\,eV has been one of the defining
X-ray spectral characteristics of the polar type CVs. However,  
the $3\sigma$ upper limit on the unabsorbed 0.3-10 keV flux of this soft blackbody, even in absence of the edge, 
is only $1\times10^{-17}$ \ergscm. We conclude that this compnent is not statistically required in the fit.
The reason for the absence of this component is discussed in Section \ref{sec:discuss}.
 
\begin{table}
\caption{Spectral parameters for the 
best fitting model.  We also report 
the absorbed/(unabsorbed) $0.3-10$ keV (F$_{0.3-10}$) flux 
and the unabsorbed $15-70$ (F$_{15-70}$), and $0.3-100$ (F$_{bol}$) keV fluxes.
Uncertainties are at the $1\sigma$ confidence level.}
\begin{center}
\begin{tabular}{ccc}
\hline 
\multicolumn{3}{c}{Average spectrum} \\
\hline 
N$_{H_{\rm P}}$             &  $10^{22}$ cm$^{-2}$ &  $0.033\pm0.003$   \\ 
N$_{H_{\rm pc}}$ 	        & $10^{22}$ cm$^{-2}$  & $4.8\pm0.5$        \\
cvf                         &  \%                  &  $42\pm2$          \\
E$_{edge}$                  & keV                  &  $0.69\pm0.01$     \\
$\tau_{edge}$               &                      &  $0.35\pm0.06$      \\
kT$_{\rm max}$              & keV                  & $37\pm7$       \\
$\alpha$                    & $1.1\pm0.1$          &                  \\
norm                        & $10^{-2}$            & $0.9\pm0.1$        \\
A$_{\rm Z}$                 &                      &  $0.71\pm0.09$     \\
EW$^{a}$                    & keV                  &  $0.10\pm0.02$     \\
F$_{0.3-10}$                &  $10^{-12}$ \ergscm  & $6.05\pm0.05$  ($\sim7.8$) \\
F$_{15-70}$                 &  $10^{-12}$ \ergscm  & $6.4\pm^{0.8}_{1.3}$  \\
F$_{bol}$                   &  $10^{-12}$ \ergscm  &  $\sim16$        \\
$\chi^2_{\nu}$ (dof)        &                      &  1.02 (1299) \\
\hline 
\\
\end{tabular}  
\label{tab:avspec}                                                                                                         
\end{center}
$^{a}$ Gaussian energy fixed at 6.4 keV.\\
\end{table}

\subsubsection{Pulse phase spectroscopy}

\begin{figure}
\centering
\includegraphics[angle=270,width=3.0in]{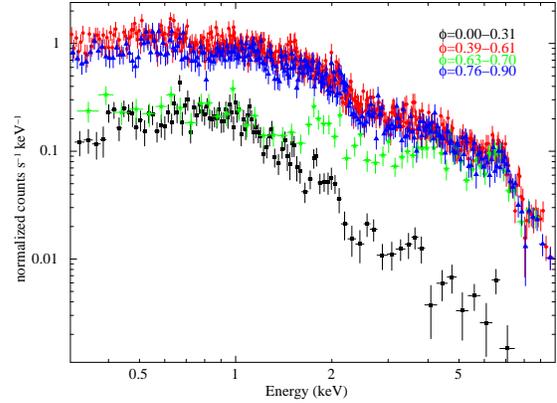}
\caption{The EPIC-PN spectra as a function of the rotational phase: minimum (black), first maximum 
(red), dip (green), second maximum (blue).  MOS1 and MOS2 spectra are not shown for plotting purposes.}
\label{fig:specphase}
\end{figure}

We also analyzed the 0.3--10 keV \XMM spectrum as a function of the rotational phase.
We selected four phase intervals corresponding to: 
the minimum ($\phi=0.00-0.31$), the first maximum ($\phi=0.39-0.61$), the dip ($\phi=0.63-0.70$), 
and the second maximum ($\phi=0.76-0.93$) of the pulse profile (see Fig. \ref{fig:x_vx_b_pulse}). 
We fitted simultaneously the spectra of the three EPIC 
cameras using the best-fitting average spectral model by linking all the model components 
with the exception of the normalisation accounting for a possible difference in the 
calibration. We fixed the interstellar absorption, 
the abundance and $\alpha$ to their average spectral value. 
All model parameters are reported in Table \ref{tab:specphase}.
We found that the spectral properties significantly change with the phase. 
The spectrum at phase minimum is characterized by a cooler emitting region (see also Fig. \ref{fig:specphase}). 
We found $kT_{\rm max}=6.3\pm0.7$, $kT_{\rm max}=55\pm10$, $kT_{\rm max}=44\pm^{50}_{16}$ and 
$kT_{\rm max}=43\pm^{12}_{8}$ keV for the minimum, first maximum, dip, and second maximum respectively. 
The N$_{H}$ of the dip is slightly higher than that of other phase intervals. 
Moreover, the covering fraction is significantly higher and it is $81\pm3\%$ 
compared to $32\pm4\%$ and $41\pm4\%$ for the first and second maximum respectively. We conclude that the dip
in the pulse profile is likely produced by an increase in the extension and density of the localized absorbing region. 
This can also explains the lower X-ray flux level for $E<1.5$ keV of the spectrum at the dip, compared 
with that of the two maxima. The spectra of three phase intervals have indeed comparable temperatures and consequently 
it is the partial covering component that produces a change in the spectral shape (see also Fig \ref{fig:specphase}).
The normalisation of the \textsc{cemekl} goes from a minimum of $0.0012\pm0.0002$ 
(pulse minimum) to a maximum of $0.0165\pm0.0003$ (first pulse maximum) following the X-ray flux trend.
We found that $\tau_{edge}$ is constant with respect to the phase within statistical uncertainty, 
while we found an indication that the equivalent width of the fluorescent Fe line could be variable. It 
goes from $0.11\pm0.03$ keV at the first maximum to $0.25\pm0.06$ keV at the dip.
We note that the addition of a blackbody component is not required in any of the phase--resolved spectra.

\begin{table*}
\caption{Spectral parameters as a function of the phase. The interstellar absorption, the abundance, and $\alpha$ are kept fix at their average spectral value:  
N$_{H_{\rm P}}=0.033\times10^{22}$ cm$^{-2}$, A$=0.71$, and $\alpha=1.1$.  
Uncertainties are at $1\sigma$ confidence level.}
\begin{center}
\begin{tabular}{cccccccccc}
\hline 
\multicolumn{10}{c}{Spectrum as a function of the rotational phase.} \\
\hline
 Phase                   & N$_{H_{\rm Pc}}$    & cvf        & E$_{edge}$    & $\tau$        & kT$_{\rm max}$    & norm           & EW$^{a}$      & F$_{0.3-10}$ $^{b}$  & $\chi^2_{\nu}$ (dof)\\
 Interval                & $10^{22}$ cm$^{-2}$ & \%          &    keV        &               &      keV          &   $10^{-2}$    &   keV         &   $10^{-12}$   &                     \\
                         &                     &             &               &               &                   &                &               &   \ergscm      &                     \\
\hline 
Min    ($\phi=0.00-0.31$)&   4.8 (fix) $^c$    &  $<50$ $^d$ & 0.70 (fix) $^c$ & $<0.51$ $^d$  & $6.3\pm0.7$       & $0.12\pm0.02$  & $<0.68$ $^d$  & $0.85\pm0.14$   & 1.26 (156)          \\
I Max  ($\phi=0.39-0.61$)& $4.0\pm0.7$         &  $32\pm4$   & $0.70\pm0.01$ & $0.50\pm0.06$ & $55\pm10$         & $1.65\pm0.03$  & $0.11\pm0.03$ & $14.0\pm0.3$    & 0.97 (747)          \\
Dip    ($\phi=0.63-0.70$)& $6.4\pm0.8$         &  $81\pm3$   & 0.70 (fix) $^c$ & $<1.30$ $^d$  & $44\pm^{50}_{16}$ & $1.29\pm0.08$  & $0.25\pm0.06$ & $11.0\pm0.7$    & 1.07 (101)           \\
II Max ($\phi=0.76-0.93$)& $3.1\pm0.6$         &  $41\pm4$   & $0.69\pm0.02$ & $0.42\pm0.08$ & $43\pm^{12}_{8}$  & $1.44\pm0.04$  & $<0.17$ $^d$  & $12.1\pm0.4$    & 1.00 (428)          \\
\hline 
\end{tabular}  
\label{tab:specphase}                                                                                                         
\end{center}
\begin{flushleft}
$^{a}$ The energy is fixed at 6.4 keV. \\
$^{b}$ Unabsorbed 0.3--10 keV flux.\\
$^{c}$ Value derived from the average spectrum.\\
$^{d}$ $3\sigma$ upper limit.\\
\end{flushleft}
\end{table*}

\subsection{The RGS spectrum}

We extracted bright phase (phase approximately 0.35--0.95) spectra (0.028$\pm$0.002 c/s, or 65\%
of total, for RGS1; 0.038$\pm$0.002 c/s, 74\%, for RGS2), with an exposure time
of $\sim19.6$ ks.  Both the phase-averaged and bright phase spectra are of
comparable statistical quality, and the results of the spectral fits are
identical, to within statistical errors.  In fitting the RGS data, we
grouped channels so that each has at least 16 source counts, used the
C statistic as the fit statistic, but used $\chi^2$ as the test statistic.

\noindent The low statistical quality of the RGS data places severe limitations
on the inferences that we can draw from them. First, we investigated
if we can confirm the presence of the absorption edge, using the best-fit
spectral model for the phase-averaged EPIC data (Table \ref{tab:avspec}),
minus the 6.4 keV line. We also compared the RGS data with this model,
but without the edge. While the addition of this component improves 
the fit, even the model without the edge is statistically acceptable. Most importantly,
from these low S/N RGS spectra, we cannot be certain that the addition of an edge, rather than
some other changes e.g., in the complex absorber, is the best way to improve the model.
Therefore, the comparison proved inconclusive.

\begin{figure}
\centering
\includegraphics[angle=0,width=3.3in]{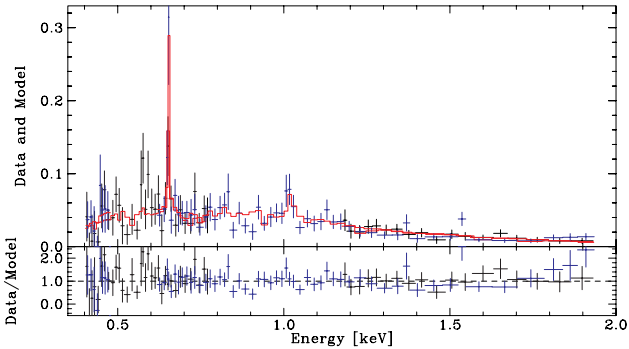}
\caption{Lightly binned RGS spectra during the bright phase. The top panel
shows the data and model (the best-fit model for the average EPIC+BAT spectrum minus the Gaussian at 6.4 keV) 
the bottom panel the data to model ratio.}
\label{RGSline}
\end{figure}

\noindent We also investigated the emission lines in the RGS spectra.  Two lines
are securely detected: the O$_{\rm VIII}$ line whose laboratory energy is 
0.654 keV ($\sim$20 eV equivalent width) and the Ne$_{\rm X}$ line expected
at 1.021 keV ($\sim$ 35 eV). In the bright phase spectra (Figure \ref{RGSline}), the Ne line has a much
lower peak, which may be because it is somewhat broadened.  Unfortunately,
at the 90\% confidence limit, only upper limits to the physical widths
can be determined ($\sigma <$ 2.6 eV for the O line and $<$19 eV for
the Ne line). The O$_{\rm VIII}$ line is measured at $0.653\pm^{0.001}_{0.002}$ 
keV and Ne$_{\rm X}$ line at at $1.013\pm^{0.009}_{0.007}$ keV. 
These best-fit line centroids are slightly redshifted compared to the laboratory 
values by $\sim450$ km/s for the O$_{\rm VIII}$ line and $\sim2000$ km/s 
for the Ne$_{\rm X}$ line, although the 90\% error ranges do (barely)
overlap with the laboratory values. Further higher S/N observations are mandatory 
to accurately estimate the redshift.
In addition, the strength of the Ne line may indicate
an overabundance of neon.  Using the best-fit EPIC model, replacing
{\tt cemekl} with {\tt cevmkl} and fixing all parameters other than the
Ne abundance and the normalisation, we obtain a Ne abundance of 2.3$\pm$1.3,
compared with 0.71 for other elements from the EPIC fit.

\subsection{Spectral energy distribution}

\begin{figure}
\includegraphics[width=6.cm,height=8.cm,angle=-90]{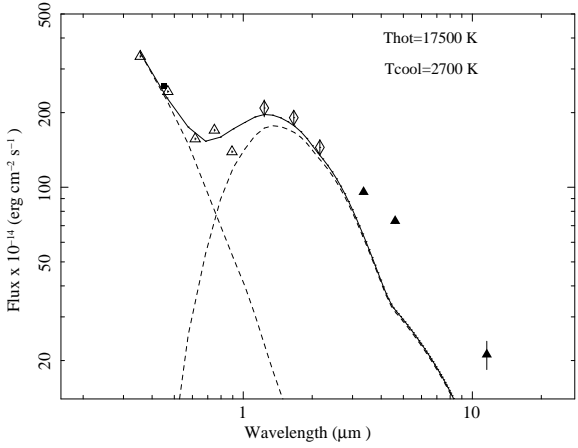}
\caption{The SED constructed with the \emph{XMM-Newton} B band 
flux (filled square), the $SDSS$ photometry (empty triangles), 
the $2MASS$ nIR (empty diamonds) and $WISE$ measures (filled triangles). 
Correction for interstellar absorption has been applied. The dashed lines 
represents two blackbody components, while the solid line is the sum of the two.}
\label{fig:SED}

\end{figure}

We also constructed a broad-band mid-IR and optical 
spectral energy distribution (SED). We used the 
average OM B band flux,  
the $SDSS$ $u',g',r',i',z'$ magnitudes, as well as the
$2MASS$ J, H and K measures and those from the   
Wide-field Infrared Survey Explorer \citep[\emph{WISE,}][]{Wright10} 
in  the W1 (3.35$\mu$m), W2 (4.6$\mu$m) and W3 (11.6$\mu$m) bands.  
We do not use the W4 (22.1$\mu$m) band as it only provides an upper limit
\footnote{The $SDSS$ data were taken in Oct. 2009, the $2MASS$ data  in 1999, 
the $WISE$ photometry in June and Dec. 2010}.
We corrected the SED for interstellar absorption 
adopting $\rm E_{B-V} = 0.05$ corresponding 
to the column density of the total absorber found from X-ray 
spectral fits ($\rm N_H = 3.3\times 10^{20}\,cm^{-2}$).
In Fig.\,\ref{fig:SED} we report the SED. It 
cannot be reproduced by a single power law as the flux
decays from the UV to the red, but increases at longer wavelengths. 
The blue portion of the SED alone cannot either be represented by 
the typical power law with index $\alpha \sim 0.5$, which is due to the combination of an
optically thick and cyclotron (free-free) emissions \citep{Harrison13}. 
We then tentatively describe the SED shape with two components consisting of a hot 
$\rm T_{h} =17500\pm 1500$\,K and a cool $\rm T_{c} = 2700\pm 300$\,K blackbodies. 
The normalisations give, for a distance of 240\,pc (see below),  
$\rm R_{h} = 6.5\pm 0.8 \times 10^{8}\,cm $  and $\rm R_{c} = 2.0\pm 0.7 \times 10^{10}\,cm $.
The former is of the order of the WD radius and could be a mixture of contributions 
(see sect.\ref{sec:discuss}).  The cooler component could be ascribed to 
the donor star. It is mainly responsible for the nIR flux $\lambda<1\mu$m, but
an excess is present at longer wavelengths ($\lambda>3\mu$m). We discuss this excess 
in Sect. \ref{sec:discuss}.

\section{Discussion}
\label{sec:discuss}

The clear detection of strong X-ray pulsations (PF consistent with 100\% in the 5-12 keV) 
at the known orbital period of \J22\ and the absence of any other periodic signal 
(e.g. from an asynchronous WD rotation period) confirms the recent proposal 
that this hard X-ray CV is a magnetic system of the polar type.  
The 2.16\,h orbital period locates this binary just at the lower edge of 
the orbital period gap and well within the range of bulk of polars \citep{beuermann99,Townsley09}.
We discuss in the following the main results of our analysis.

\subsection{The orbital variability} 

\noindent The X-ray light curve folded at the orbital period reveals the typical
bright and faint phases of the magnetic polar-type CVs.
These are due to the post-shock accretion flow above the main (upper) pole coming into (bright phase) 
and out of view (faint phase) if the magnetic axis is tilted with respect to the rotation 
axis ($\beta$) and shifted in azimuth ($\psi$).
Here $\beta$ is the magnetic colatitude and $\psi$ is the angle between the line of 
centres of stellar components and the projection of the magnetic axis on the orbital
plane \citep[see][]{Cropper88}. The length of the faint phase is used to 
restrict the range of binary inclination $i$ and $\beta$ \citep{Cropper90}.
The presence of one bright phase implies $i + \beta \geq 90^o$.  The absence of eclipses implies $i\leq 75^o$. 
\cite{Thorstensen13}, adopting $\rm M_{WD} = 0.75 \rm M_{\odot}$ and $\rm M_{2} = 0.2 \rm M_{\odot}$
for the WD and companion masses, suggest $i =50^o$. We extend the orbital inclination to $i =42^o$,
as the WD mass is found to be larger (see below). The faint phase lasts $\Delta \phi \sim$0.3, giving
$\beta \sim 55^o - 64^o$  for $i \sim 40^o - 50^o$. Furthermore, the rapid rise and decay of the bright
phase can be used to estimate the extent of the X-ray emitting region. 
They both last $\sim$500\,s which would imply an azimuthal extent of the 
spot on the WD of $\sim 24^o$  assuming no lateral extent.
This translates into a fractional circular spot area $\sim$0.045\,$\rm A_{WD}$, 
where A$_{WD}$ is the total WD area. 

\noindent The X-ray bright phase is characterised by a narrow (width$\sim0.1$)  dip that
is energy dependent due to absorbing material above the X-ray emitting region.
A dip due to the accretion stream at the threading of the magnetic field lines would imply $i > \beta$, but
we find it is not the case. The only absorbing material that would 
produce such a feature 
should be located in the accretion pre-shock flow \citep{done_and_magdziarz98}. 
This feature is also observed at hard X-rays $>5$\,keV. At these
energies it cannot be due to absorption. It has been observed in a few other poalrs and 
explained as due to occultation by the dense core of the channeled accretion flow \citep{ramsay04a}. 
The shifts observed in the time occurrence of the dip during the \XMM\ observation also
suggest changes in the local mass accretion rate. This is also seen in other few polars
\citep{HarropAllin99,Schwope01}.
Due to the lack of eclipses and the impossibility to link the spectroscopic ephemeris by
\cite{Thorstensen13} with  our epoch, we cannot constrain the azimuth $\psi$ of the
accreting region.

\noindent The $0.3-12$ keV X-ray light curve during the faint phase does not reach zero counts. 
This implies that there is an additional contribution, likely 
the secondary pole, that comes into view as the WD rotates. 
The spectrum of the faint phase indicates that this pole
has a softer emission. This is not unusual in the polars, where the main pole is harder than
the other one \citep{beuermann99}. 

\noindent The B band light curve also displays a bright phase, which is almost
in phase with the X-rays as well as the dip feature, although the rise leads the X-rays by $\sim700$ s.
The optical emission in polars can be due to several contributions, such
as cyclotron emission, the accretion stream, the accretion heated WD spot as well as the unheated
WD photosphere. These
are difficult to disentangle without polarimetric and multi-colour photometric data. Cyclotron beaming 
is expected to be anti-phased with respect to the X-rays since the maximum cyclotron flux is emitted 
perpendicularly to the field lines \citep{Cropper88,Cropper90,Matt00,Burwitz98}.
An upper limit to the contribution of cyclotron flux can be derived from the flux 
difference at maximum and minimum of  optical light curve. If the variability is totally
due to cyclotron, we derive $\rm F_{cyc}/F_{X} < 0.11 $, indicating that cyclotron is not 
the main cooling mechanism in \J22\ . 
The B band light curve has a smoother rise and lower amplitude than the X-rays, which suggests
a larger emitting area and being due to reprocessing \citep[see also][]{Vogel08}. The 
broad-band SED indicates a mildly hot component contributing to the blue 
portion. This comprises of  the variable and stationary components, which we cannot disentangle due
to the lack of  phase--resolved colour photometry. 
The presence of the dip in the
optical with similar width as that in the X-rays, further indicates that the dense parts of 
the accretion flow are responsible for this feature also in the optical. 
Polarimetry and colour-photometry are then essential to identify the origin of the optical
variability.

\subsection{The spectral characteristics}
\label{subs:spec_car}

\noindent The X-ray analysis reveals an average hard X-ray spectrum 
with maximum temperature 
$kT=37$ keV. This is consistent with the typical temperature of magnetically confined accretion flows, 
where a shock region is formed above the poles of the WD. kT$_{max}$ must be regarded as a lower 
limit of the maximum temperature of the post-shock region, 
which strongly depends on the WD mass \citep{aizu73}. We also note that
the presence of the Fe line at 6.4\,keV would require a reflection component
 above 10\,keV \citep{Done95}. Its inclusion would lower 
the maximum temperature. However, due to the low S/N of the data above 10\,keV,
this component is not required in the fit.
Furthermore, the lack of a detectable soft blackbody component and the fact that this
polar is a strong hard X-ray emitter with respect to other polars 
(see also below) allows us to use the X-ray temperature to estimate
the WD mass. In particular we use the dedicated model 
of \cite{suleimanov05}\footnote{This is aprivate code that runs
into \textsc{Xspec}} developed for bremsstrahlung dominated magnetic CVs and 
in particular for IPs. 
The model accounts for the growth of pressure towards the WD surface and hence the change of gravity. 
This allow to get a more reliable estimate of the maximum
temperature and, consequently, a more solid estimate of the WD mass. 
The model is computed for the continuum only, consequently we added a
broad Gaussian to account for the iron complex (thermal and
fluorescence) centred at $6.68\pm0.03$ keV and
with EW$=0.53\pm0.07$ keV. A fit to the \XMM\ EPIC and Swift BAT spectrum 
for E$>3$ keV gives $M=0.99\pm^{0.09}_{0.14}\,M_{\odot}$, but the fit quality 
is slightly low ($\chi^{2}_{\nu}=1.15$). 
As a cross-check, we also derived the WD mass from the temperature 
of the spectrum at the first maximum, $kT_{max}=55\pm10$ keV, and we get 
$\rm M_{wd}=0.97\pm0.08$ M$_{\odot}$ \citep[see][for more details]{aizu73}. 
This is fully consistent with what derived from the dedicated model. Such a WD would have a radius 
of $\rm R_{wd}=0.57\pm0.06\times 10^{9}$ cm.

\noindent The spectral analysis also show the presence of high column density material 
($\sim3-6\times10^{22}$ cm$^{-2}$), which  partially absorbs  the X-ray emission. This material is located in the 
pre-shock flow above the WD pole and is the main responsible in producing the pulse dip.

\noindent There is an indication that the EW of the fluorescent Fe line at 6.4 keV could changes with the phase, 
being slightly larger in the dip than at the maximum of the pulse profile. 
This could indicates that the Fe line is less sensitive to orbital 
variations than the continuum. Whether it forms at the WD surface or/and
in the pre-shock flow cannot be assessed without knowledge of a reflection
component at hard X-rays, which is expected to be present along the fluorescent
iron line \citep{Done95}.

\noindent Particularly interesting is the detection in the EPIC spectra of ionized absorbing material 
that produces an absorption feature at $\sim0.7$ keV. 
This energy  might be consistent with an absorption edge of O$_{\rm VII}$. Unfortunately, we
cannot confirm this feature in the RGS due to the poor S/N.
An oxygen edge has been observed in a few IPs \citep[][]{Mukai01,demartino08,Bernardini12}
and commonly in the
low mass X-ray binaries dippers. If due to oxygen \J22\ would be the first polar
showing such feature. 
The ionization state in the pre-shock flow depends on the ratio of matter
density in the pre-shock to the radiation density emitted in the post-shock regions. A 
low level of ionization is expected if cyclotron,
which is emitted in a non-ionizing form, is the dominant cooling mechanism or if
the shock is taller than wide. In tall shocks a large fraction of 
X-rays are emitted perpendicularly to the field lines.
Hence, both the evidence of a warm absorber and the indication 
of a large extent of the accretion spot in \J22\ would corroborate 
a high ionization state in the pre-shock flow. This in turn would 
also suggest that cyclotron is not the dominant cooling mechanism, 
likely due to a low magnetic field WD.

\begin{figure}
\includegraphics[width=\columnwidth,height=6.5cm,angle=0]{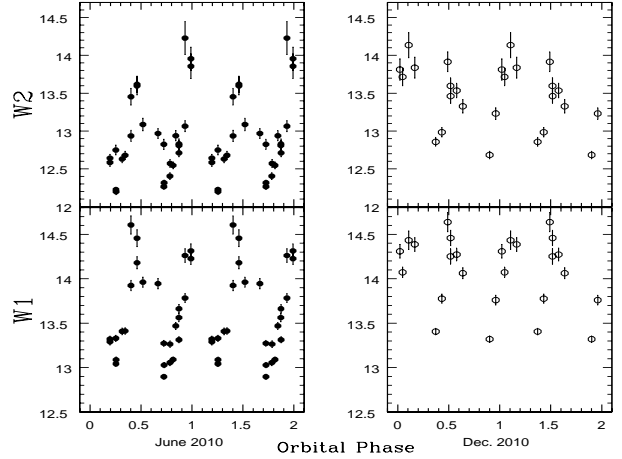}
\caption{The \emph{WISE} light curve constructed using 
the "single exposures" in the W1 and W2 bandpasses for the two epochs 
(June and Dec. 2010) showing large IR  variability. The orbital phases are computed 
using arbitrary zero times. Ordinates are in magnitudes.}
\label{fig:WISE_lc}
\end{figure}

\noindent The SED also reveals 
a cool ($\rm T_{c}\sim2700$ K) emission from the companion star. 
We note an excess of flux in the \emph{WISE} bands for $\lambda>3\mu$m. 
Another argument favouring an extra mid-IR component is that 
time-resolved \emph{WISE} photometry (see Fig.\,\ref{fig:WISE_lc}) 
shows a large variability $\Delta m \sim$1.6 
that cannot be due to ellipsoidal modulation of the donor star 
(typically less than 0.1\,mag). However, caution has to be taken, 
because \emph{WISE} and $2MASS$ data are taken 10 years apart.
A mid-IR excess is questioned to be due
to dust by circumstellar discs or by cyclotron from the accretion region.  
A circumbinary disc would not give rise to any variability and, therefore,
the mid-IR excess is very likely due to cyclotron emission. 
Similar behaviour was found for AM\,Her and EF\,Eri 
and the few polars observed in the mid-IR \citep{Brinkworth07,Harrison13},
indicating that low cyclotron harmonics ($n= 1,2,3$), at these wavelengths, are optically thin. 
Both polars share a a low magnetic field strength (B$\sim1.3\times10^{7}$ G). 
The lack of nIR and IR spectroscopy does not allow a meaningful analysis 
with cyclotron models for \J22.

\noindent For an orbital period of 2.16\,h, the secondary star is expected 
to be of spectral type M4\,V, $\rm T_{eff}$ = 3300\,K \citep{knigge06}. 
The temperature we derived from the SED is compatible within 2$\sigma$ 
with this value. The $2MASS$ (J-H) colour corrected for interstellar extinction is 0.70$\pm$ 0.11\,mag.
This is consistent, within uncertainties, with that expected for a M4\,V star, 
$\rm (J-H)_{o}$=0.57 \citep{knigge06,Straizys2009}. Using a M4\,V star 
absolute magnitudes $\rm M_J = 8.45$ and $\rm M_K = 7.60$ \citep{knigge06}  
and assuming that it contributes 82$\%$ and 89$\%$ to the observed flux
in these bands, as derived from the SED, 
we estimate a distance of 230-250\,pc, respectively.


\subsection{The energy balance}

We now estimate the mass accretion rate and  
verify if it is consistent with that expected for 
a binary evolving through gravitational radiation.
For a donor star mass $\rm M_{2} = 0.2 \,M_{\odot}$, the binary mass ratio is $q=0.2$ and
for $\rm P_{orb}$=2.16\,h the expected secular mass transfer rate for gravitational losses \citep{Warner95} is 
$\rm \dot M \sim 4 \times 10^{-11}\, M_{\odot}/yr$. 
At a distance of 230-250\,pc the bolometric
X-ray luminosity is $\rm L_X = 1.1\pm0.1\times10^{32}\,erg\,s^{-1}$. Assuming
it represents the accretion luminosity, for a WD mass 0.97\,M$_{\odot}$,
we obtain that the mass accretion rate is $\rm \dot M \gtrsim 8 \times 10^{-12}\,M_{\odot}/yr$. 
We conclude that, despite the uncertainties on the true total accretion luminosity, 
the mass accretion rate is consistent with a binary evolving through gravitational radiation.

\noindent The lack of a soft (20--50\,eV) blackbody component in the X-ray spectrum, 
could arise by the large accretion footprints that would shift 
the reprocessed radiation to EUV, FUV wavelengths.
\emph{XMM-Newton} observations have shown that a large
fraction  ($\sim 40\%$) of polars when accreting at high rate do not exhibit
a detectable soft blackbody component \citep{ramsay04b}.
This suggests selection effects in source detection due to soft band coverage of previous 
X-ray surveys \citep[e.g. ROSAT][]{beuermann99}. 
 \J22\ thus adds to this group. However, as pointed out by \citep{Ramsay09} there is no obvious
explanation on why these hard polars should have larger accretion areas than 
the other polars since neither the orbital period nor magnetic field strengths of the 
two groups are so different.

\noindent The hard X-ray detection of \J22\ also poses the question on whether the few 
hard X-ray selected polars possess lower magnetic fields and/or more massive WDs. To date 11 
systems have been detected in the \Swift\ and \INT\ surveys with a few asynchronous ones. 
These do not seem to share similar magnetic field strengths and masses (de Martino, in 
prep.). \J22\ is found to have a moderately massive WD. Whether its magnetic field is 
low, it should be assessed with future polarimetric observations.

\section{Summary}
\label{sec:summary}

We studied \J22, a candidate CV of the polar type.
The main goal of this research is to unambiguously 
unveil the source real nature and characterize 
its broad-band properties. For this purpose we used  
X-ray and optical \XMM\ data, together with non simultaneous 
high-energy coverage provided by \Swift\ BAT and ground based 
optical, near and mid-IR data. The main results are briefly summarized in the following.

\begin{itemize}

\item We measure a X-ray period of 2.16 h, fully consistent 
with that derived from optical spectroscopy, thus allowing us to 
identify \J22\ as a CV of the polar type.

\item The source shows a longer bright and a shorter faint phase with a prominent 
dip between the two, both in the X-ray and in the optical band. 
The pulse shape changes with the energy, and the dip is more pronounced 
at lower energy. The pulsed fraction strongly increases with the energy 
interval being consistent with $100\%$ at high energies. These are all 
typical characteristics of polars.

\item We infer an orbital inclination $i \sim 40-50^o$, a magnetic colatitude
$\beta \sim 55-64^o$ and a large accretion spot azimuth $\sim 24^o$.

\item  The X-ray emission originates in the post-shock flow reaching a maximum
temperature of $\sim$ 55\,keV and is heavily absorbed by high density 
material in the pre-shock flow. We derive a WD mass of 0.97$\pm$0.08\, M$_{\odot}$. 
The lack of a soft X-ray blackbody component adds \J22\ to the 
increasing group of polars without a detectable reprocessed X-ray component.

\item An intense fluorescent Fe line at 6.4 keV is present whose
 equivalent width changes along the orbital cycle. 

\item An absorption edge at 0.70\,keV possibly due to 
to O$_{\rm VII}$, reveals for the first time the presence of a warm absorber in a polar.

\item We found indication of an overabundance of neon and a possible indication of
redshift of the O$_{\rm VIII}$ and Ne$_{\rm X}$ lines of about 450--2000 km/s.

\item The optical to mid-IR SED is described with two blackbody components at $\sim$18000\,K
and at $\sim$ 2700\,K. While the former could be a mixture of contributions difficult
to assess with the present data,  the cool component is ascribed to the donor star of M4\,V spectral type.
A mid-IR excess also suggests the presence of  an extra component, possibly 
due to  emission from lower cyclotron harmonics.
We also estimate a distance of 230-250 pc.

\end{itemize}

\section*{Acknowledgments}
This work is based on observations obtained with
\XMM\ an ESA science mission with instruments and contributions
directly funded by ESA Member States; with \Swift, a NASA science
mission with Italian participation. This publication also makes use of data products 
from the Wide-field Infrared Survey Explorer, which is a joint project of the University of California, 
Los Angeles, and the Jet Propulsion Laboratory/California Institute of Technology, 
funded by the National Aeronautics and Space Administration; the Two Micron All Sky Survey ($2MASS$), 
a joint project of the University of Massachusetts and the Infrared Processing and Analysis Center (IPAC)/Caltech,
funded by NASA and the NSF; and the Sloan Digital Sky Survey ($SDSS$).\\ 
FB acknowledge the Galileo Galilei Institute for Theoretical Physics for the hospitality and the INFN for 
partial support during the completion of this work, and thank Dr. Giorgio Lanzuisi for his precious 
suggestions concerning Monte Carlo simulations. DdM and FB acknowledge financial support by ASI under contract ASI-INAF I/037/12/0.

\bibliographystyle{mn2e}
\bibliography{biblio}

\vfill\eject
\end{document}